\documentclass[aps,prl,reprint,nofootinbib]{revtex4-2}
\usepackage{blindtext}
\usepackage{amsmath}
\usepackage{graphicx}
\usepackage{float}
\usepackage{comment}
\usepackage{xcolor}
\usepackage{ulem}
\usepackage{mathrsfs}

\newcommand{\mE}{\mathcal{E}}
\newcommand{\sE}{\mathcal{U}}

\def\mnras{\ref@jnl{MNRAS}} 
\def\pasj{\ref@jnl{PASJ}}  

\begin{document}
\title{Experimental observation of violent relaxation and the formation of out-of-equilibrium quasi-stationary states.}
\author{M.~Lovisetto$^1$,  M.C. Braidotti$^{2}$, R.~Prizia$^{2,3}$, C.~Michel$^4$, D.~Clamond$^1$,  M.~Bellec$^4$,  E.M. Wright$^{5}$, B.~Marcos$^1$,  D. Faccio$^{2,3}$}
\affiliation{$^1$ Universit\'e C\^ote d'Azur, CNRS, Laboratoire J.-A. Dieudonn\'e, 06109 - Nice, France}
\affiliation{$^2$ School of Physics \& Astronomy, University of Glasgow, G12 8QQ Glasgow, UK}
\affiliation{$^3$ Institute of Photonics and Quantum Sciences, Heriot-Watt University, EH14 4AS Edinburgh, United Kingdom}
\affiliation{$^4$ Universit\'e C\^ote d'Azur, CNRS, INPHYNI, France}
\affiliation{$^5$ Wyant College of Optical Sciences, University of Arizona, Tucson, Arizona 85721, USA}

\begin{abstract}
Large scale structures in the Universe, ranging from globular clusters to entire galaxies, are the manifestation of relaxation to out-of-equilibrium states that are not described by standard statistical mechanics at equilibrium. Instead, they are formed through a process of a very different nature, i.e. {\it violent relaxation}. However, astrophysical time-scales are so large that it is not possible to directly observe these relaxation dynamics and therefore verify the details of the violent relaxation process. We develop a table-top experiment and model that allows us to directly observe effects such as mixing of phase space, and violent relaxation, leading to the formation of a table-top analogue of a galaxy. The experiment allows us to control a range of parameters, including the nonlocal (gravitational) interaction strength and quantum effects, thus providing an effective test-bed for gravitational models that cannot otherwise be directly studied in experimental settings.
\end{abstract}

\maketitle

\textit{Introduction} ---  The observable Universe is populated with objects and structures that evolve over time whereas galaxies and globular clusters appear to be macroscopically stationary objects at thermodynamic equilibrium \cite{Binney2006}. However, Chandrasekhar pointed out in 1941 that the time necessary for these objects to reach thermal equilibrium is actually much larger than their age \cite{Chandrasekhar1941}. 
This has been confirmed  by observations determining that  these astrophysical structures are indeed far from thermal equilibrium (see, e.g., \cite{Anguiano_2020}). 
Lynden-Bell in 1967 proposed a mechanism, {\it violent relaxation}, by which these out-of-equilibrium quasi-stationary structures can actually form \cite{Lynden-Bell1967}. It has been subsequently understood that the formation of these states is generic in Hamiltonian systems  with a long range interacting potential, i.e., a potential that is not integrable as a result of its extension over large 
scales \cite{Campa2009}. This phenomenon is similar to what arises in plasmas subject to 
Landau damping, in which there is an exchange of energy between the electromagnetic wave 
generated by the particles of the plasma and the particles themselves \cite{Landau1946}. 
Landau damping has been observed in plasma experiments \cite{Malmberg1964, Neil1965,Laslett1965,Damm1970,Gentle1971,Sugawa1988,Danielson2004} and in space plasma 
turbulence \cite{Chen2019}. \\
Violent relaxation, however, is more elusive and has not been observed to date in a repeatable or controllable experiment. Indeed, experimental observation of the dynamics of the formation of quasi-stationary states via violent relaxation is hindered  for essentially two reasons. First, there are systems in which it is potentially present, but it is destroyed by the stochastic noise generally present in these systems \cite{Chalony2013}. Second, there are systems in which violent relaxation is actually present, but the associated timescales are too large to actually be observed. This is the case of astrophysical systems such as galaxies, independently if it is constituted by classical (non-quantum) dark matter particles (see e.g. \cite{zwicky_1933,boyarski_2009,sofue_2009,cupani_2010}), or  composed by quantum matter (see e.g. \cite{Wayne2000,HsiYu2014, Hui2017,Marsh2018, Alexander2019}). 
In these systems violent relaxation occurs in timescales of the order of million years \cite{Binney2006}.\\
Here, using a table-top nonlinear optics experiment, we report the experimental observation of a violent relaxation process and the subsequent formation of a quasi-stationary state in the form of a ``table-top galaxy''. The ability to also tune the parameters of the interaction provides a valid test-bed to compare theory and observations, and a new approach to the study of the dynamics of long range interacting systems.\\

\textit{{Self-gravitating systems.}}
The temporal evolution of self-gravitating particles of dark matter, of mass $m$, defined by a wavefunction $\psi$, is described in 3D by the Newton--Schrödinger equations (NSE):
\begin{subequations}
\label{SNA}
\begin{align}
&\mathrm{i} \hbar \partial_t \psi + \frac{\hbar^2}{2 m}\nabla^2 \psi + m \phi \psi=0 \\
&\nabla^2 \phi = -4\pi G |\psi|^2,
\end{align}
\end{subequations}
where $|\psi|^2$ is the mass density, $G$ the gravitational constant and $\nabla^2$ the three-dimensional (3D) Laplacian.
The gravitational potential, $\phi$, generated by the mass distribution itself, depends on the constant $G$ and the mass density.
When the system is in the semi-classical regime, which corresponds to $\hbar/m\ll 1$, the main processes leading towards a quasi-stationary state are usually characterized by two distinct phenomena \cite{Binney2006}: mixing and violent relaxation. The former, which is caused by the evolution of the density distribution in the gravitational potential, mixes the phase-space while conserving the distribution of energy density.
The latter consists in the evolution of the distribution of energy as a result of oscillations of the potential. Mixing alone can give rise to a quasi-stationary state, but violent relaxation makes the process much more efficient.
In this case, the quasi-stationary solution corresponds to the formation of an oscillating soliton in the center of the system (defined as the ground state of Eq.~\eqref{SNA}, see e.g. \cite{Moroz1998})  surrounded by a classical solution, usually described by the Vlasov-Poisson equation, which is the $\hbar/m\to0$ limit of the NSE \cite{HsiYu2014,Mocz2018}. {The characteristic size of the soliton, $\xi$,  can be estimated by calculating the scale for which the kinetic and potential energies are of the same order of magnitude, giving $\xi=\hbar/m \sqrt{s/8\pi GM}$, where $M$ is the total mass of the system and $s$ the characteristic size of the whole system. To monitor the degree of classicality, we define the parameter $\chi=\xi/s$, the semi-classical limit corresponding to $\chi\to 0$.}\\

\textit{{Optical system.}} 
The evolution of the amplitude, $\mathcal{E}$, of a monochromatic laser beam propagating through a thermally focusing nonlinear medium in the paraxial approximation is described by \cite{Roger2016, Bekenstein2015, Navarrete2017}:
\begin{subequations}
\begin{align}
&\mathrm{i}\/ \partial_z \mathcal{E} + \frac{1}{2 k_0 n_b}\nabla_\perp^{\,2}\/ \mathcal{E} 
+ k_0 \Delta n \mathcal{E}=0, \\
&\nabla_\perp^{\,2}\/ \Delta n = -\frac{\alpha \beta}{\kappa} |\mathcal{E}|^2.
\end{align}
\label{SNO}
\end{subequations}
The operator $\nabla_\perp^{\,2}$ is the transverse two-dimensional (2D) Laplacian, $k_0=\frac{2 \pi}{\lambda}$ the wave-number of the incident laser with $n_b$ the background refractive index of the medium. The non-local nonlinear refractive index change, $\Delta n$, is induced by the beam itself heating the medium.
$\beta$ is the medium thermo-optic coefficient, $\kappa$  its thermal conductivity and $\alpha$ its absorption coefficient.\\
Provided that the propagation axis, $z$, plays the role of time, $t$, the similarity between Eqs.~(\ref{SNA}) and (\ref{SNO}) underpins the opportunity to directly observe 2D violent relaxation in a laboratory experiment.\\
{We define a transverse length scale for which both the linear and nonlinear effects are of the same order as $\xi = \sqrt{z_{\rm nl}/(2k_0 n_b)}$. $z_{\rm nl} = \kappa/(\alpha \beta k_0 P)$ is the longitudinal length over which the effect of the nonlinear term becomes substantial and $P$ is the power of the laser beam. Preparing the initial beam with transverse width $s$ dictates the propagation regime of the system. For $\chi = \xi /s = 1$, a soliton of characteristic size $\xi$ is expected to form. The optical equivalent of the above-mentioned semi-classical regime is obtained when $\chi \ll 1$. Although $\chi$ is evolving with $z$, this specific condition is maintained during the propagation.}\\
Moreover, one can define a local energy density of the optical system as
\begin{equation}
\sE(\mathbf{r}, z)=\frac{\left|\nabla_\perp \mE(\mathbf{r},z)\right|^2}{2k \left| \mE(\mathbf{r},z)\right|^2} - k_0\Delta n(\mathbf{r},z).
\label{energy}
\end{equation}
The first contribution corresponds to the kinetic (linear) energy, the second one to the potential (nonlinear) energy. The total energy is a conserved quantity.
In order to characterise and quantify mixing and violent relaxation in optical experiments, we define two quantities.
First, the Wigner transform \cite{Wigner1932} $F(\boldsymbol{r},\boldsymbol{k},z)$ of the optical field $\mathcal{E}$ is 
the density of probability to find a portion of the optical beam at the position $\boldsymbol{r}$ with wavevector $\boldsymbol{k}$. 
We use the evolution of $F$ with respect of $z$  to study the  mixing 
of the phase-space. Second, the evolution of the distribution of energy density $\nu(\sE)$ of the optical field $\mathcal{E}$, captures the main signatures of the violent relaxation process (see Methods for detailed expressions).\\

{\textit{Experimental setup.} 
Figure~\ref{setup}(a) shows a schematic representation the experiment. A continuous-wave laser beam with a Gaussian profile and wavelength $\lambda = 532$ nm propagates in a thermo-optical nonlinear medium made of three aligned identical slabs of lead-doped glass for a total length $L=30$ cm, represented here as a single slab.
\begin{figure*}
\centering
\includegraphics[width=2\columnwidth]{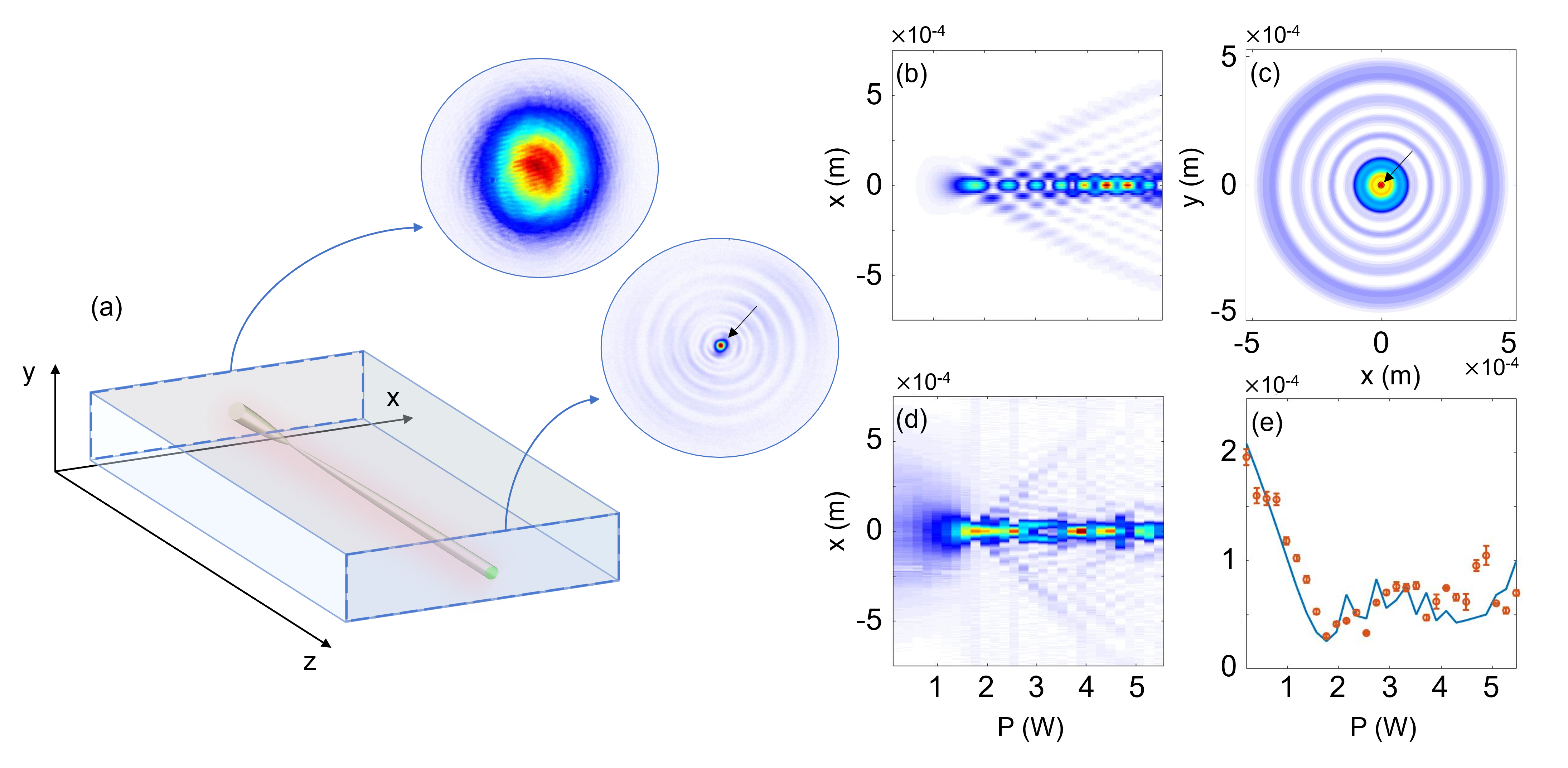}
\caption{\label{setup} (a) Sketch of the experiment: a Gaussian laser beam propagating in a lead-doped crystal. The diffusion of heat inside the nonlinear medium is represented by the glowing red profile. Insets show the input and output experimental profiles at $P=5$ W. (b),(d) $y=0$ slice of the beam intensity profile as a function of one transverse coordinate $x$ and power, obtained from the numerical simulation (b) and experimental data (d). (c) Simulation of the transverse plane $(x,y)$ at $z=L$ for the power $P=5$ W. We observe the soliton (red dot indicated by the black arrow) surrounded by the classical part corresponding to $\chi\to0$. (e) Comparison between experiment (red dots with error-bars) and simulation (blue curve) of the average size of the beam profile as a function of power.}
\end{figure*}
The beam width $s=350~\mu$m  at the sample input facet is chosen experimentally by a system of lenses (not shown) such that the condition $\chi \ll1$ is fulfilled (see details in Methods). When the intense laser beam propagates inside the crystal, it induces a nonlocal interaction (heating) of the medium (depicted as a red glow around the propagating beam in Fig.~\ref{setup}(a)). The beam at the output facet of the medium is imaged onto a camera, where we
collect its interference with a reference beam (not shown). By using the Off-Axis Digital Holography (OADH) technique \cite{bertolotti}, we are able to access the the spatial distribution of both the intensity and the phase of output field. To explore the full dynamics of the laser beam, we tune the initial power from 0.2 W to 5.5 W. The insets in Fig.~\ref{setup}(a) show the experimental beam intensity profile at the input and output crystal facets for input power $P=5$ W. The output intensity profile shows the expected central soliton (ground state, indicated in the figure), surrounded by the classical solution. \\ 
{Experimentally, it is only possible to access the field at the output facet of the sample and not at the full nonlinear propagation inside the material. However, by expressing the propagation coordinate in terms of the relevant dynamical characteristic scale $z_{\mathrm{dyn}}=s\sqrt{n_b\kappa /\alpha\beta P}$ (see Methods), one can show that varying the initial power $P$ and measuring the intensity at fixed $z = L$ as a function of $L/z_{\mathrm{dyn}}$ is equivalent to measuring the intensity at different steps $z$ inside the material at fixed $P$ as a function of $z/z_{\mathrm{dyn}}$ (now with $z_{\mathrm{dyn}}$ fixed). There is hence a direct mapping between power $P$ and propagation length $z$, when $\chi \ll 1$. Tuning the power, $P$, we are able to follow the $z$-evolution of the beam amplitude $\mathcal{E}$ corresponding to the time-evolution of the mass distribution in astrophysics. We therefore hereafer use $P$ to parameterize the system evolution [see Figs.~\ref{setup}(b-e)]. 
}
}\\

{\textit{Collapse and quasi-stationary state of the system.}
Figure~\ref{setup}(b) and (d) depict the intensity profiles (along $y=0$) measured at the output of the glass sample as a function of power $P$ obtained from the numerical simulations (details in Supplementary Discussion) and experiments, respectively. We observe good qualitative agreement of two main features, i.e., the initial collapse that is then followed by a stabilization showing that for large $P$, the system is reaching a quasi-stationary state. A plot of the simulated intensity distribution for the power $P=5$ W  is shown in Fig.~\ref{setup}(c). It illustrates the typical expected solution which combines a solitonic part (the high intensity peak in the center of the structure) with a surrounding classical part (compare to the experimental inset in Panel (a)). A quantitative comparison is provided in Fig.~\ref{setup}(d) by plotting the average size of the beam $R$ (see Methods) for both the numerical simulation (curve) and experiments (circles). We observe a good agreement and the small differences  in the oscillatory part after the collapse  can be expected due to their chaotic nature \cite{Vandervoort2011}, which therefore strongly depend on the experimental input conditions.
}\\

\begin{figure*}
\centering
\includegraphics[width=2\columnwidth]{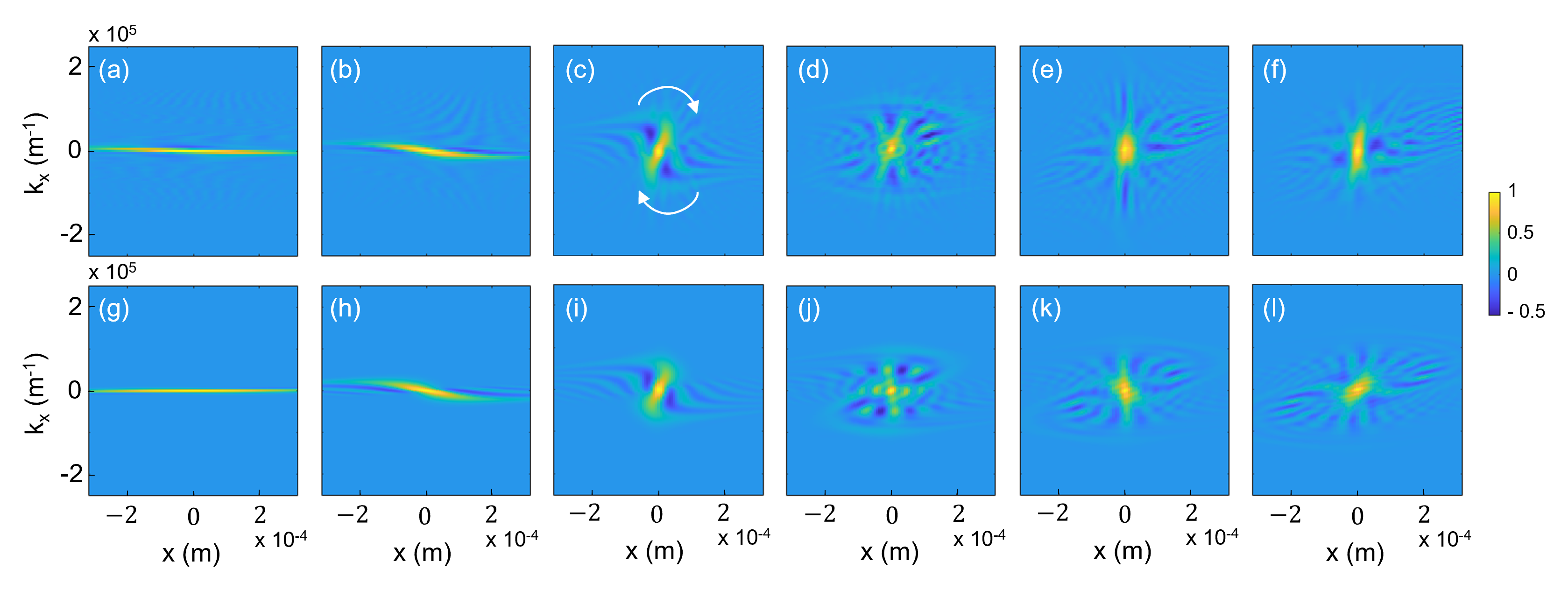}
	\caption{Results of experiment - top row (a)-(f) - and simulation - bottom row (g)-(l) -  for the $y = 0, k_y = 0$ profiles of the Wigner distribution at different powers: (a),(g) $P= 0.2$W, (b),(h) $P= 1$W,(c),(i) $P= 2$W,(d),(j) $P= 3$W,(e),(k) $P= 4$W,(f),(l) $P= 5$W. }
	\label{husimi_comparison}
\end{figure*}

{\textit{Phase-space mixing.}
We next study the existence of mixing in the system by analysing the evolution of the phase-space. To this end, we use the Wigner distribution (see Methods) of the full complex-valued optical field. The experimental and numerical results are presented in Fig.~\ref{husimi_comparison} and, here again, are in a good agreement. At the initial stage, the system has a Gaussian spatial distribution with a very narrow dispersion along the $k_x$-axis. As $P$ increases, the phase mixing starts by first rolling up the phase-space (indicated by the  white arrows) and then forming characteristic filaments \cite{Binney2006}. In the Supplementary Discussion, we show numerically that in a system where only mixing is present, such as in the Snyder-Mitchell model \cite{snyder1997accessible}, the evolution of the system is significantly different. 
In the following, we demonstrate an additional and direct signature of violent relaxation by studying the evolution of the distribution of energy density of the system.
}\\

 \medskip

\begin{figure}
\centering
\includegraphics[width=\columnwidth]{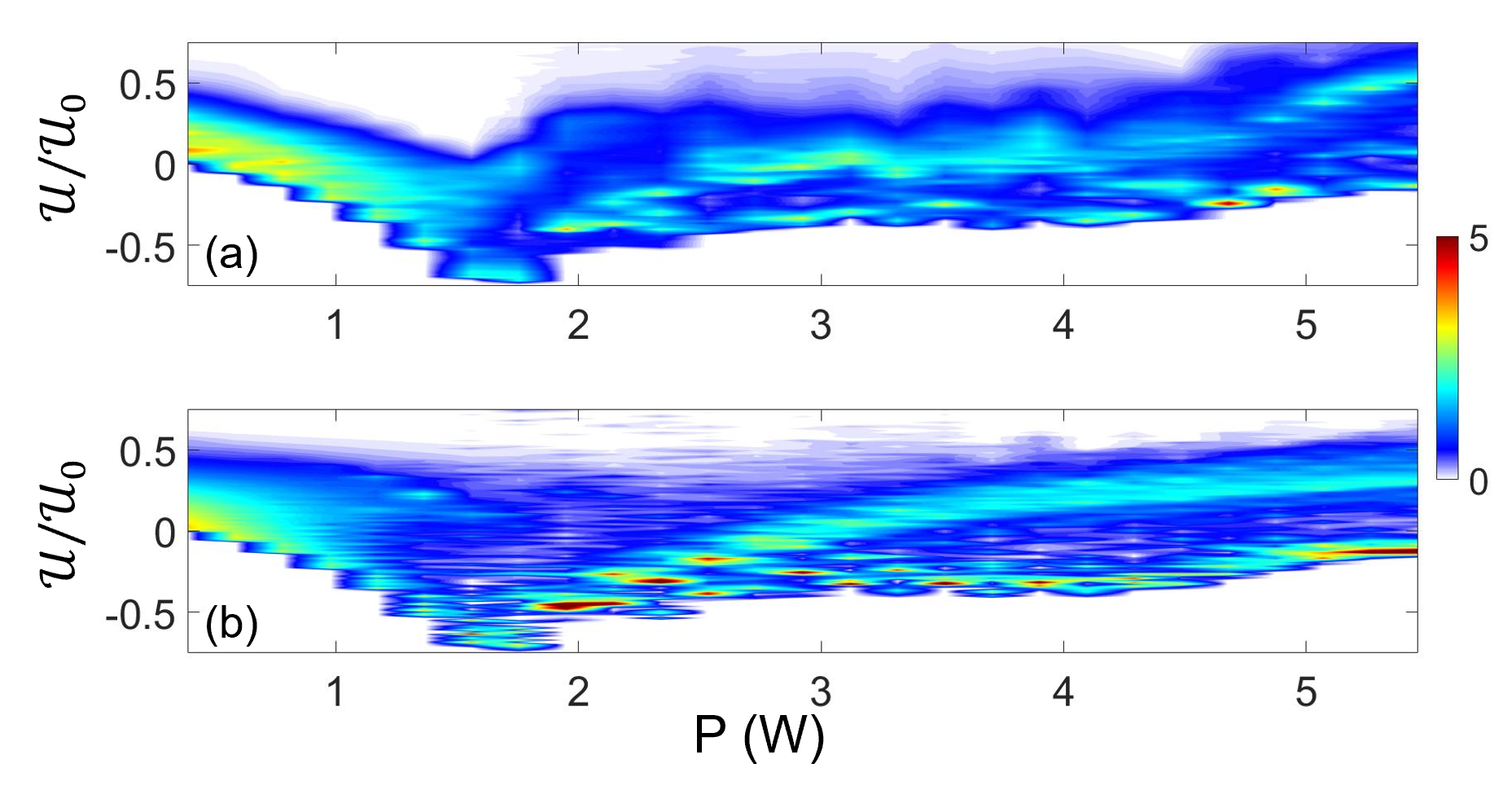}\\
	\caption{Distribution of energy density map a for the experiment (a) and simulation (b).  
}
\label{energy_comparison}
\end{figure}

{\textit{Direct observation of violent relaxation.}
In a classical system (i.e., $\chi \to 0$) with no losses, the only mechanism responsible for a change in the distribution of energy density is the violent relaxation process \cite{Binney2006}. 
Figure~\ref{energy_comparison} shows the experimental (a) and numerical (b) distribution of energy density, $\nu(\mE)$, obtained for various input powers, $P$. 
Before the minimum collapse (around $P\approx 1.8$ W), we observe that the distribution of energy density globally decreases because it is dominated by the potential energy and the system is collapsing.  In contrast, 
after the collapse, the distribution of energy density exhibits two characteristic `structures', which persist for the whole subsequent evolution: one at smaller energies, which corresponds to the inner region which has already completely relaxed. A second `structure' at higher energies is related to the more peripheral regions, which have not completely relaxed yet.
We observe however, as power increases after the collapse, the distribution of energy density tends asymptotically to a quasi-stationary state. At higher powers ($>5$ W) we observe (more clearly in the simulations) also the formation of a soliton that is associated with the minimal energy of the system.}\\

{\textit{Discussion.}}
These results highlight the details of the violent relaxation dynamics. These do rely on the condition $\chi \ll 1$, so as to isolate the classical dynamics that we are interested in here. In our experiments, {$\chi = 2.3 \times 10^{-2}/\sqrt{P}$, with $P$ measured in Watts (see Methods) and is therefore of order $10^{-2}\ll1$ over the full evolution. However, at the same time the effect of the finite value of $\chi$ is actually still visible in the Wigner distribution (Fig.~\ref{husimi_comparison}) where the negative regions correspond to quantum effects (see e.g. \cite{Heller1977}).  Equation~\eqref{SNO} gives the Heisenberg uncertainty relation $\Delta k_x \Delta x\sim 1$, which corresponds to the typical size of the observed negative regions. These can be seen  to be relatively small compared to the total surface in which the Wigner function is non-zero, indeed as a direct consequence of $\chi\ll 1$. Moreover, it is possible to explicitly show that $\chi$ is also sufficiently small as to have negligible effects also on the distribution of energy density (Supplementary Discussion). }\\
We also note that the experimental system intrinsically exhibits losses (required to induce the nonlocal interaction through weak absorption of the beam). The total loss (including also air-glass interface reflections) is estimated to  be $\sim50\%$. These losses modify the second term of Eq.~\eqref{energy}. 
However, we verified through numerical simulations (see Supplementary Discussion), that the presence of losses in the experiment has a limited impact on the distribution of energy density. Furthermore, the field evolution is only {weakly}  modified. Drawing a connection with astrophysics, absorption and losses would correspond to a loss of mass in the system that does not alter the global dynamics and the presence of violent relaxation.\\

\begin{figure}
\centering
\includegraphics[width=0.7\columnwidth]{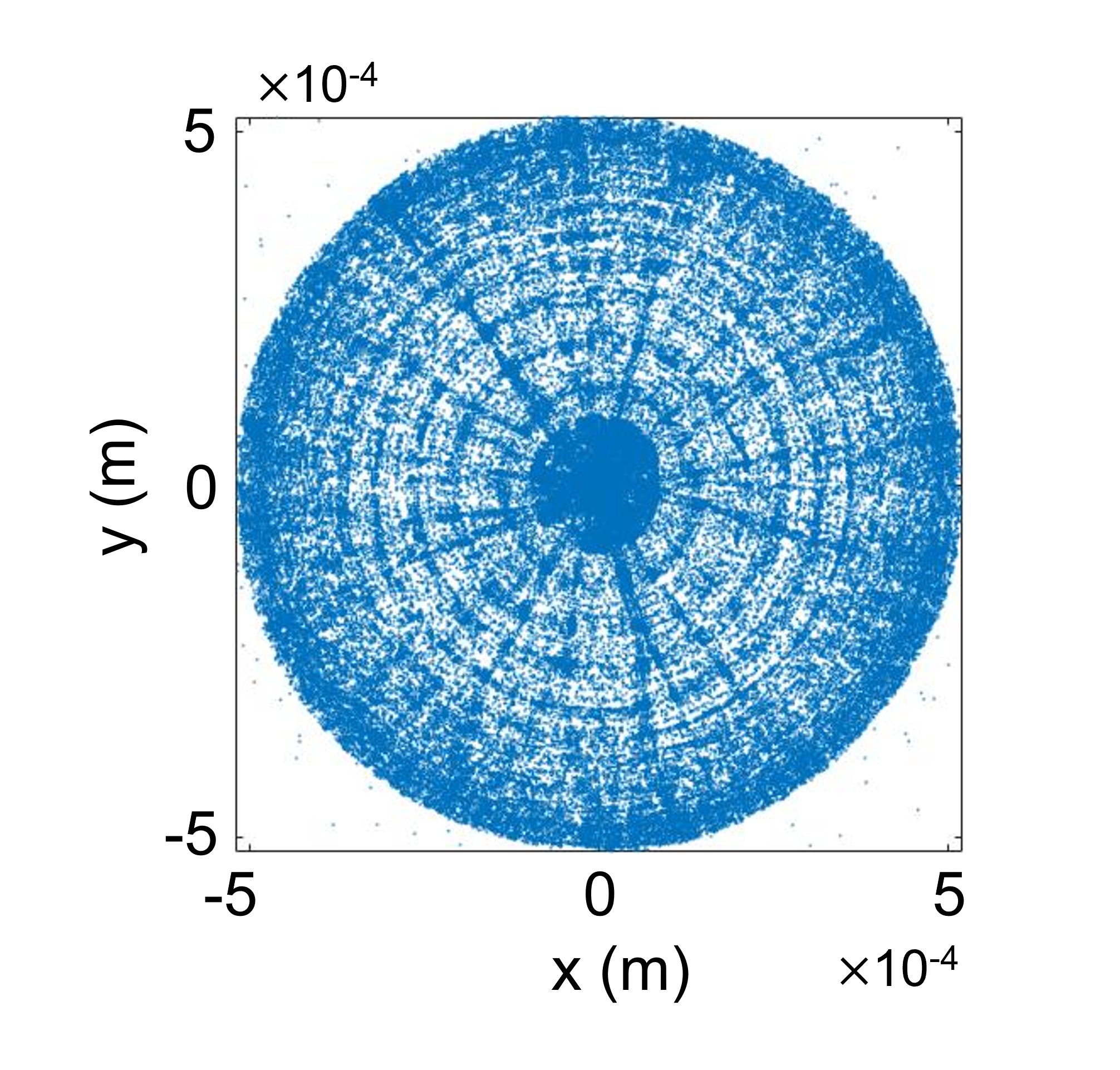}
\caption{\label{particles} N-body simulation performed evolving self-gravitating particles of dark matter. We have used $2^{17}$ N-body particles and the parameters map directly on to those used in the experiments (see SM for details), i.e. this galaxy is the particle version of the optical galaxies shown in Fig.~\ref{setup}(a) inset (experiment) and (c) (numerical simulation).  }
\end{figure}

{\textit{Conclusions.}} We have provided experimental evidence of  violent relaxation in a long interaction-range system that gives a direct confirmation of the formation of an out-of-equilibrium stationary state that follows the scenario advanced by Lynden-Bell in 1967 \cite{Lynden-Bell1967}. Our experiments relate to observable galaxies  whose formation dynamics are not directly observable or at least, are not repeatable.  With our table-top experiments, we can directly connect our parameters to those of a particle-based system, as shown in Fig. \ref{particles}, corresponding to the  galaxy distribution for a particle system with parameters equivalent to those of the experiment, to be compared with Fig.~\ref{setup}(c), corresponding to numerics (see also SM). The parameters can then also be tuned in order to make the system more or less classical, i.e., to tune up to which spatial scale quantum effects are important.
Here, we focused on the classical evolution, where the spatial scale of analogue quantum effects is one order of magnitude smaller than the size of the system. \\
%
%
The  next steps may cover further aspects of long range systems such as investigating the effect of angular momentum, studying mergers of structures (which are known as the main mechanism of the formation of spiral galaxies), and simulating systems corresponding to various Dark Matter models. \\
\section{Methods}
\small{
{\textit{Experiment}} --- The experimental setup is shown in Fig.\ref{Exp_Setup2}. A continuous-wave laser with a Gaussian profile with wavelength $\lambda = 532$ nm is split into 2 components: a reference beam and a target beam. The reference beam is expanded using a system of lenses and incident onto a CMOS camera. The target beam is shaped to have waist $s=350~\mu$m (waist calculated where the intensity falls of $1/e^2$ - the value has been obtained by a Gaussian fit of the beam intensity at the sample input face - see inset in Fig. \ref{setup}) and shines onto three aligned identical slabs of lead-doped glass (height $D = 5$ mm, width $W = 40$ mm and length $L_0 = 100$ mm each, hence a total length $L=300$ mm), represented as a single crystal. \\
The glass is a self-focusing nonlinear optical medium with thermal conductivity $\kappa = 0.7$ Wm$^{-1}$K$^{-1}$, background refractive index $n_b = 1.8$, absorption coefficient $\alpha = 1 m^{-1}$, thermo-optic coefficient $\beta = \frac{\partial n}{\partial T} = 2.2 \cdot 10^{-5}$ K$^{-1}$ and transmission coefficient at the sample interface $T=0.92$. The value of the coefficient $\beta$ is found by a fit of the experimental beam evolution and results to be 1.6 times larger than the value provided by the manufacturer. With these experimental parameters, we have $z_{\mathrm{nl}}\approx 2.7/P$~mm and $\chi = 2.3 \times 10^{-2}/\sqrt{P}$. As explained in the main text, since it is only possible to measure $\mE$ at the end of the sample, in order to explore its value inside the sample we make use of mapping between propagation distance $z$ and power. This mapping holds if the parameter $\chi$ is kept constant and therefore it is necessary to vary the width $s$ of the initial condition (see the definition of $z_{dyn}$ and $\chi$ in the main paper). However, as shown in the Supplementary Discussion, for sufficiently small values of $\chi$, which is the case in our setup, the experiment is weakly sensitive to a variation of $\chi$. Therefore, we keep $s$ constant when varying $P$ to simplify the experimental procedure.\\
\begin{figure}[b]
\centering
\includegraphics[width=8.5cm]{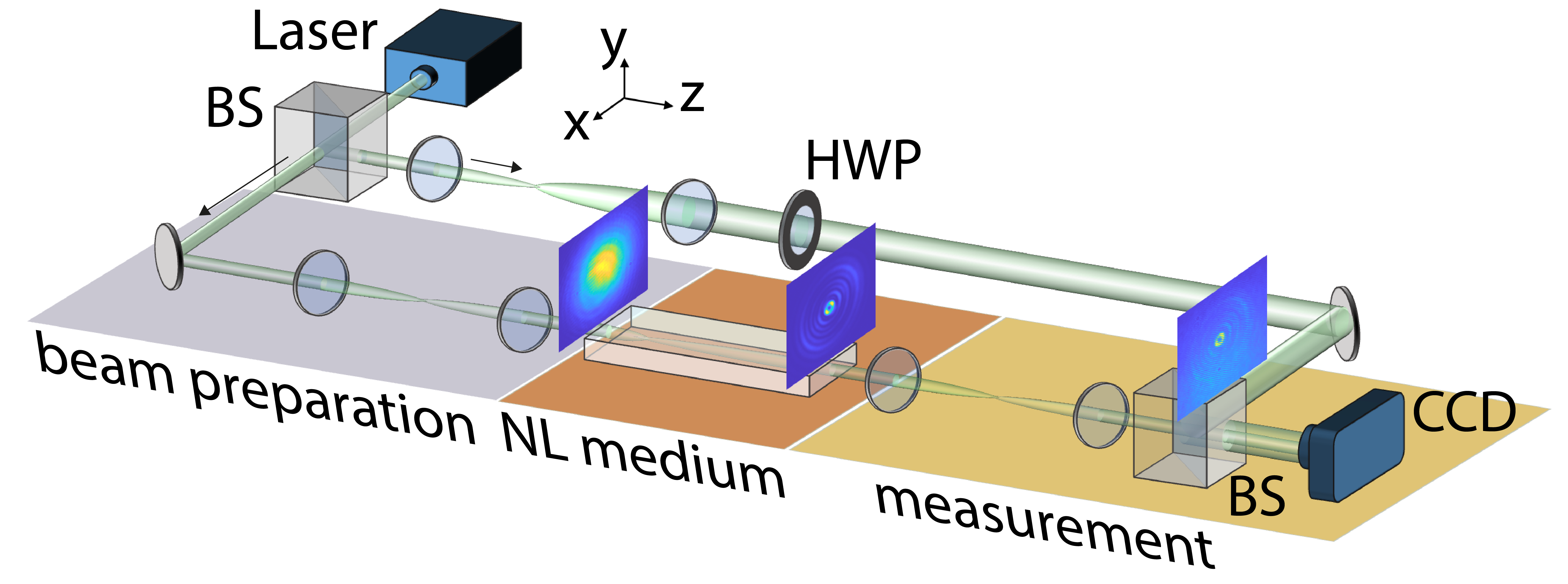}
\caption{\label{Exp_Setup2}  Experimental setup: a monochromatic laser beam is split into 2 components: a target and a reference
beam. The reference beam is expanded using a system of lenses and incident onto a CMOS camera. The target beam is imaged onto the front face of the nonlinear sample through a 4f-imaging system that also allow to choose the input beam waist size. 
After the nonlinear propagation, the target beam at the output sample face is imaged onto a CMOS camera, where we collect its interference with the reference beam (all beams have the same optical frequency $\lambda = 532$ nm). Half-wave plates ($HWP \lambda/2$) are placed along the beams paths in order to finely tune the beam
polarizations in order to have the maximum visibility of the interference fringes. We collect with the CMOS camera
the interferograms of the mixed field and the reference. By means of the off-axis digital holography technique \cite{bertolotti}, we are able to reconstruct the amplitudes and phases of the target beam. Target beam input powers range from $0.2 \,W$ to $5.5 \,W$, with a $0.25 \,W$ step. Top images in the set-up sketch show the experimental intensity profile of the beam at that point in propagation distance: first image from the left is the input beam intensity, then there is the output facet beam profile ($P=5$ W); last image is the interference of the output target with the reference beam. }
\end{figure}
\textit{Data analysis} ---  The experimental intensity profiles are characterized by a background noise - this is removed by averaging out the intensity at pixels that are at the edge of the $(x,y)$ beam profiles; this average is used as an estimate for the background noise and then subtracted from the whole experimental data. We then apply a noise mask, i.e. the intensity points far from the main body of the beam profile are set to zero. On the other hand, the interferograms do not need the noise removal, since the off axis digital holography technique requires a Fourier-transform of the beam which automatically filters all high-frequency contributions from the signal.\\
\textit{z to P mapping} --- A natural dynamical characteristic length scale $z_{\mathrm{dyn}}$ appears in the regime $\chi\ll1$. This can be calculated writing the corresponding Newton equation of Eq.~\eqref{SNO}: $n_b k_0 d^2 \vec r_\perp /dz^2 = k_0 \nabla_\perp  \Delta n$, where $\vec r_\perp$ is the position in the transverse plane. Using that the typical size $r_\perp\propto s$ and hence $\nabla_\perp\/ \Delta n \propto \Delta n /s$ and the initial velocities $d \vec r_\perp /dz\simeq 0$, we get $z_{\mathrm{dyn}}=s\sqrt{n_b\kappa /\alpha\beta P}$. This expression allows to map $P$ with $z$.\\
\textit{Observables} --- The size of the system is measured using the quantity $R(z)=(P_{y_0})^{-1} \int I(x,y=0,z) \left|x\right| d x$
 with $P_{y_0}=\int I(x,y=0,z) dx$. 
 
 Using the polar symmetry of the beam amplitude, we compute the Wigner transform  \cite{Wigner1932}  on the $(x,k_x)$ plane as
\begin{equation}
    F(x, k_x,z)=\int dx'\, \mE\left(x+x'/2,0, z\right) \mE^*\left(x-x'/2,0, z\right)e^{i k_x x'},
    \label{wigner}
\end{equation}
where $F(x, k_x,z)$ is a representation of the classical density of probability to find a piece of beam at the position $(x,0)$ with wavevector $(k_x,0)$. 
The distribution of energy density is defined as
\begin{equation}
\nu(\sE)=\frac{1}{P}\int d^2r\, \delta\left[\sE-\sE(\mathbf{r},z) \right]I(\mathbf{r},z),
\label{energy-distribution}
\end{equation}
where $\delta$  is the Dirac delta function. 
}


\small{
{\textit{Acknowledgements.}} The authors acknowledge financial support from EPSRC (UK Grant No. EP/P006078/2) and the European Union’s Horizon 2020 research and innovation program, Grant Agreement No. 820392. D.F. acknowledges financial support from the Royal Academy of Engineering Chair in Emerging Technology programme. M.L. and B.M. acknowledges support by the grant Segal ANR-19-CE31-0017 of the French Agence Nationale de la Recherche.\\

{\textit{Author contributions.}} B.M. and D.F. conceived the project and ideas. M.L., M.C.B. and R.P. performed the experiments, data analysis and numerical simulations. B.M. and E.M.W. performed theoretical analysis. D.C. contributed to numerical simulations. C.M. and M.B. contributed to data analysis.  All authors contributed to the manuscript.

}

\end{document}


\title{Supplementary Information}
\date{}
\maketitle

\section{Numerical Methods}

\subsection{Newton--Schr{\"o}dinger equation}

The numerical scheme employed to solve the Newton--Schr{\"o}dinger equation is a 
Split-Step algorithm for the propagation along $z$, 
combined with a pseudo-spectral method for the integration 
over the transverse $(x,y)$-plane. Since in our configuration 
the boundaries are sufficiently far from the laser beam, we 
observe a very weak dependence of the simulation results with 
respect to the boundary conditions used (Fig.~1).
Therefore, for simplicity in numerical and analytical calculations, 
boundary conditions are taken into account by means of the so-called 
Distribution Loss Model \cite{Roger2016}, where a degree of 
non-locality $\sigma = D/2$ is introduced to describe the diffusion 
of heat in the system. As initial condition for the numerical simulations, we use a fit of the input beam injected into the sample, i.e., the  Gaussian field %
\begin{equation}\label{ICgauss}
    \mathcal{E}_{\mathrm{fit}}(x,y,z=0)\ =\ A\, \ue^{-\frac{(x-x_0)^2+(y-y_0)^2}{2s^2}}\,\ue^{-\ui k_0 \frac{(x-x_0)^2+(y-y_0)^2}{2f}},
\end{equation}
where the parameters resulted to be $s=350 \,\mu\mathsf{m}$ and 
an initial phase $f=-1\,\mathsf{m}.$

Absorption in the material is taken into account introducing an 
extra term in Eq.~(2a) of the main paper:
\begin{align}
\ui \,\frac{\partial \mathcal{E}}{\partial z}\ +\ \frac{1}{2\, k}\, \nabla_{\!\perp}^{\ 2}\,\mathcal{E}\ +\ k_0 \,\Delta n\,\mathcal{E}\ +\ \ui\,\frac{\alpha}{2}\,\mathcal{E}\ =\ 0, 
\end{align}
where $\alpha = 1\,\mathsf{m}^{-1}$ for the material used in the 
experiment. In addition, reflections at the interfaces of the three 
aligned samples are taken into account introducing a drop in the 
total intensity by a factor $1-T$, where $T=0.92$ is the 
transmittance at each interface. 
\begin{figure}[]
\begin{center}
	\includegraphics[width=1\linewidth]{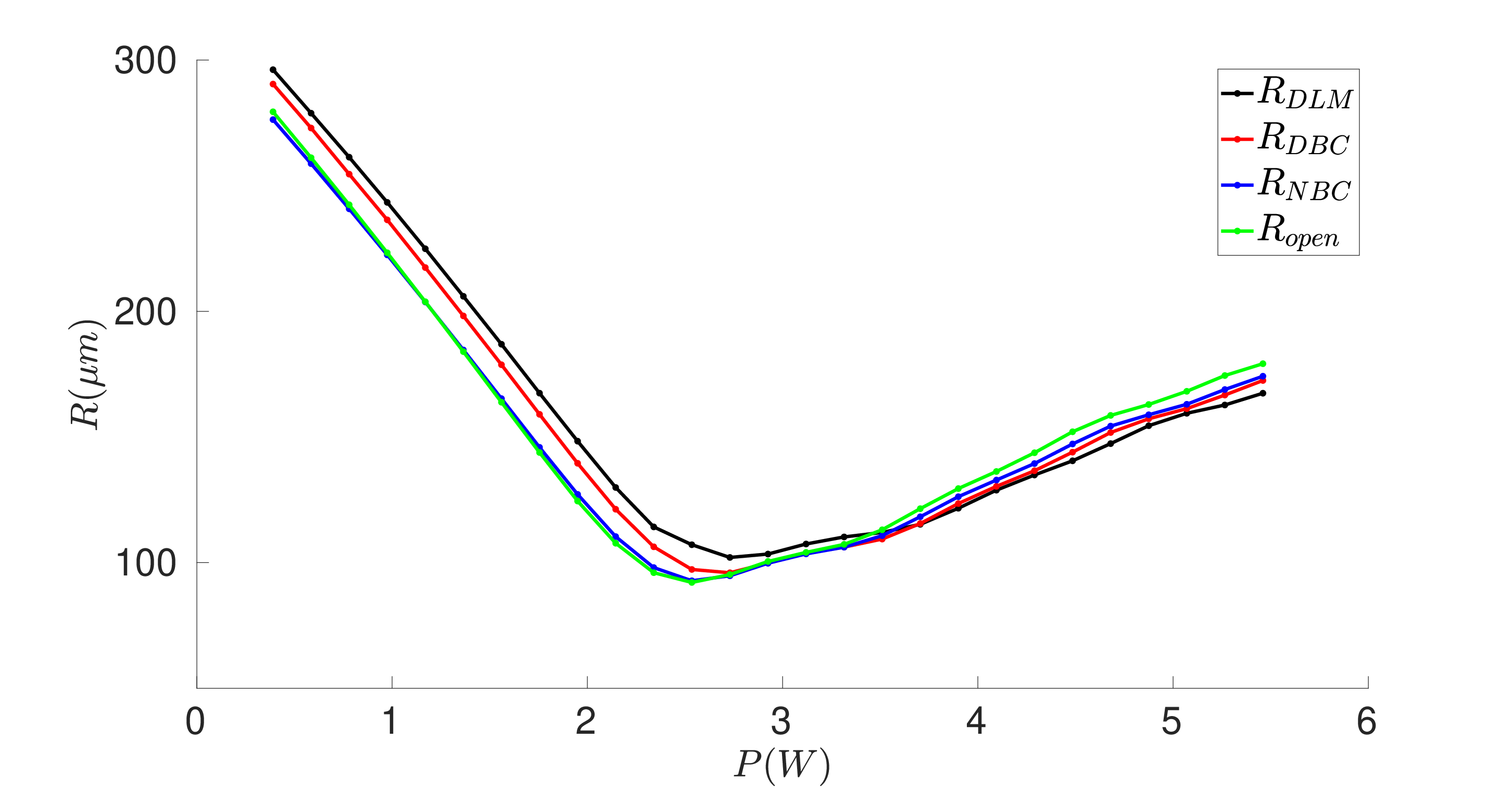}
	\caption{Comparison between the outputs of the Newton--Schr{\"o}dinger equation simulation with the Distribution Loss Model (DLM), Dirichlet (DBC), Neumann (NBC) and open boundary conditions (open) for the average size of the beam profile as a function of power at the end of the sample.}
	\end{center}
	\label{DLMvsDBC}
\end{figure}

\subsection{N-body method}

As explained in the main paper, in the $\hbar/m \to 0$ limit the Newton--Schr{\"o}dinger equation is equivalent to the classical evolution, given by the Vlasov-Poisson equation. However, the computational cost needed to numerically solve the latter equation is extremely large; for this reason $N$-Body methods are usually employed \cite{hockney2021computer}. 
In this case the mass distribution function, $f(\boldsymbol{r},\boldsymbol{v},t)$ is sampled with a set of $N$ bodies. The sampling procedure is analogue to a Monte-Carlo method.
The resulting equations are
\begin{equation}\label{Nbodyeq}
    k\,\frac{d^2\,\boldsymbol{r}_i}{dz^2}=\frac{\alpha\,\beta\,P}{2\,\pi\,\kappa\,n_b}
\sum_{j=1}^{N}\frac{(\boldsymbol{r}_i-\boldsymbol{r}_j)}{(\left|\boldsymbol{r}_i-\boldsymbol{r}_j\right|^2+\epsilon^2)^{3/2}},
\end{equation}
where $\boldsymbol{r}_i=(x_i,y_i)$, $P$ is the power and $\epsilon$ is a smoothing parameter. The latter removes the singularities in the potential and suppresses small-scale fluctuations, which are due to the discrete nature of the $N$-body approach \cite{hockney2021computer}. Notice that, applying the mapping $z \mapsto t$, $k \mapsto m$ and $(\alpha\,\beta\,P)/(2\,\pi\,\kappa\,n_b) \mapsto G\,m$, \eqref{Nbodyeq} corresponds to the classical equation of motion for a system of $N$ self-gravitating particles, all with the same mass $m_i=m$. However, the $N$ bodies must be interpreted as tracers of the phase-space distribution rather than actual particles.

The initial condition is generated performing a Poisson sampling process of the Gaussian appearing \eqref{ICgauss}, while velocities are initialized with the gradient of the initial phase of the complex exponential in \eqref{ICgauss}.

Concerning losses, reflections are taken into account in the same way as in the NSE case, while absorption in the material is introduced multiplying the right hand side of \eqref{Nbodyeq} by $\ue^{-\alpha\,z}$. The smoothing parameter is set to $\epsilon=10^{-3}\,s$.

\section{Discrimination of the evolution of phase space and density with or without violent relaxation} 

The mechanism of the mixing process is driven by the existence of a 
(static) an-harmonic potential. The violent relaxation  mechanism,  
on the other hand, requires the presence of a $z$-dependence on the 
potential. In this section, we show that looking at the evolution of 
the NSE model (which presents phase-space mixing and violent relaxation) 
and  the Snyder--Mitchell  model \cite{snyder1997accessible} (which 
presents only phase mixing) leads to different evolutions which 
gives a strong indication of the existence of violent relaxation in 
the system. The SM model consists in a Schr\"odinger equation coupled 
with a nonlinear refractive-index profile calculated  using the {\it 
input} beam profile $\mathcal{E}_0=\mathcal{E}(\vec r_{\perp},z=0)$, 
namely:
\begin{equation}
\begin{gathered}
\ui\,\partial_z\,\mathcal{E}\, +\, \frac{1}{2 k}\,\nabla_{\!\perp}
^{\ 2}\, \mathcal{E}\, +\, k_0\, \Delta n\, \mathcal{E}\, =\,0, \\
\nabla_{\!\perp}^{\ 2}\,\Delta n\, =\, -\frac{\alpha \beta}{\kappa}
\,I_0,
\end{gathered}
\end{equation}
where $I_0=|\mathcal{E}_0|^2$.

\begin{figure}[H]
\minipage{0.56\textwidth}
\resizebox{\textwidth}{!}{\includegraphics[width=\linewidth]{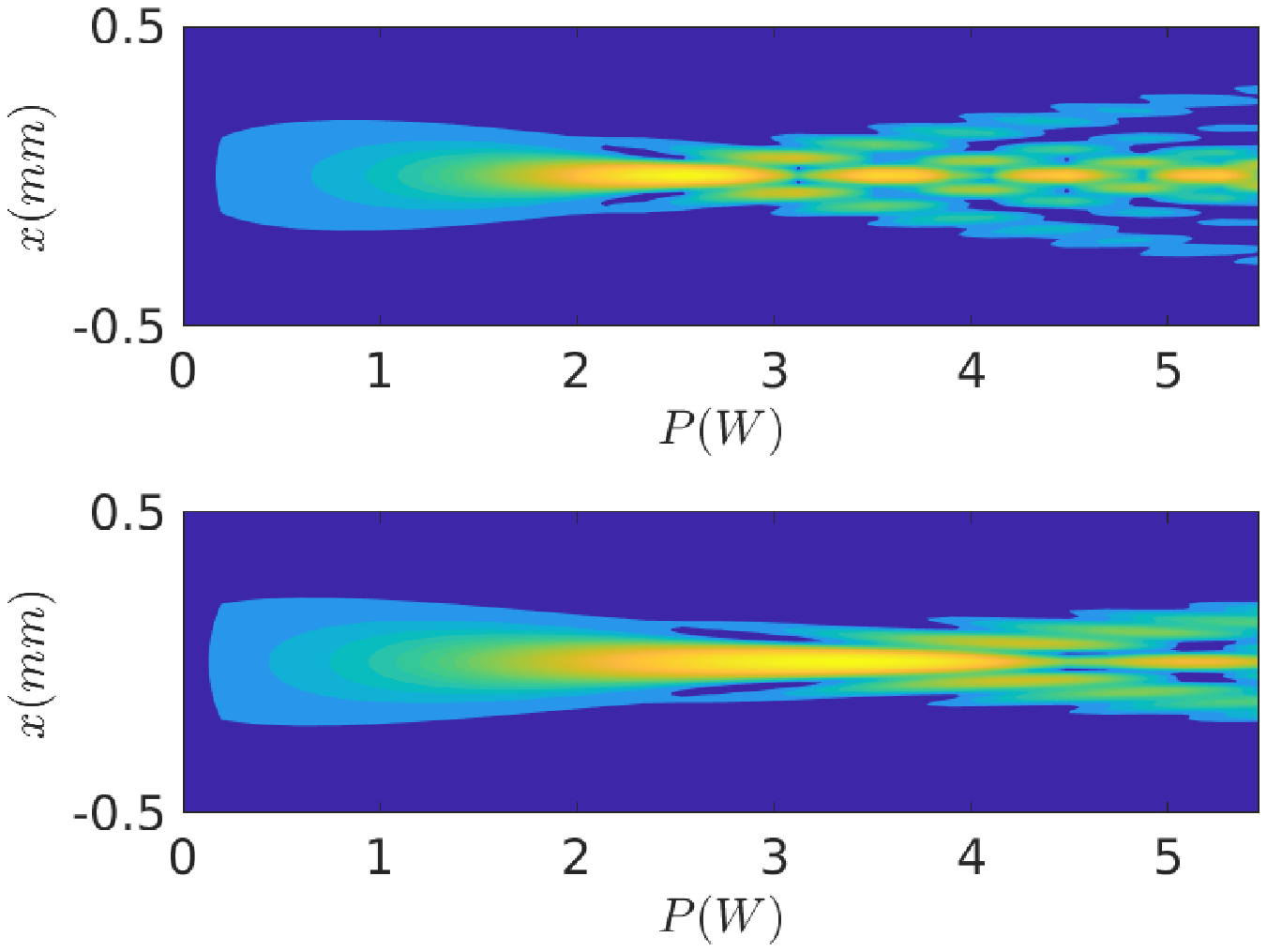}}
\endminipage\hfill
\minipage{0.4\textwidth}
\resizebox{\textwidth}{!}{\includegraphics[width=\linewidth]{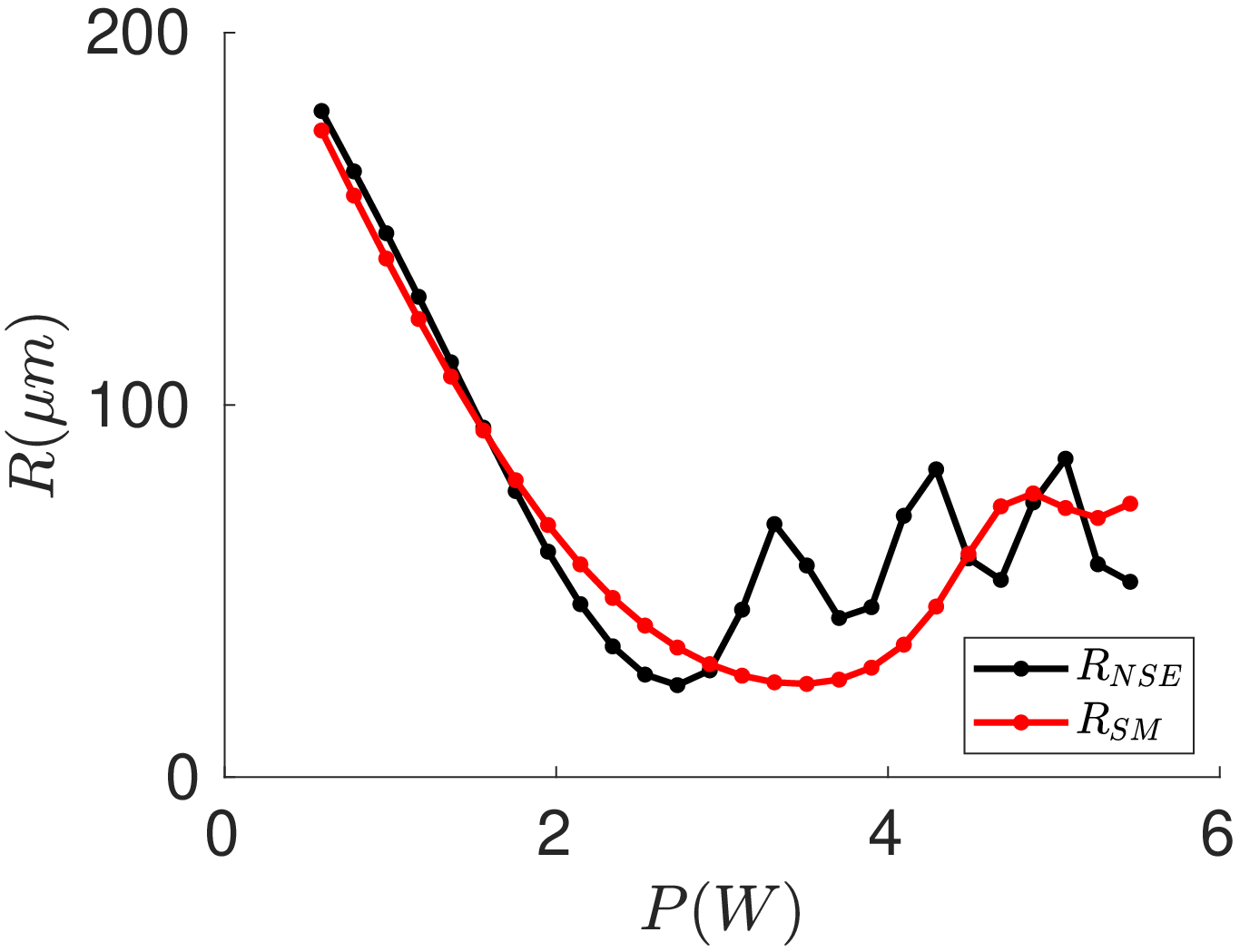}}
\endminipage
	
\caption{Left plots: $y=0$ slice of the beam intensity profile as a function of one transverse coordinate $x$ and power, obtained from the NSE (top left) and SM model (bottom left); both plots are in logarithmic color-scale. Right plot: comparison between the outputs of the NSE simulation (black curve) and SM model (red curve) for the one dimensional average size of the beam profile as a function of power.}
\label{NSEvsSM1}
\end{figure}

\begin{figure}[H]
\centering
\includegraphics[scale=0.8]{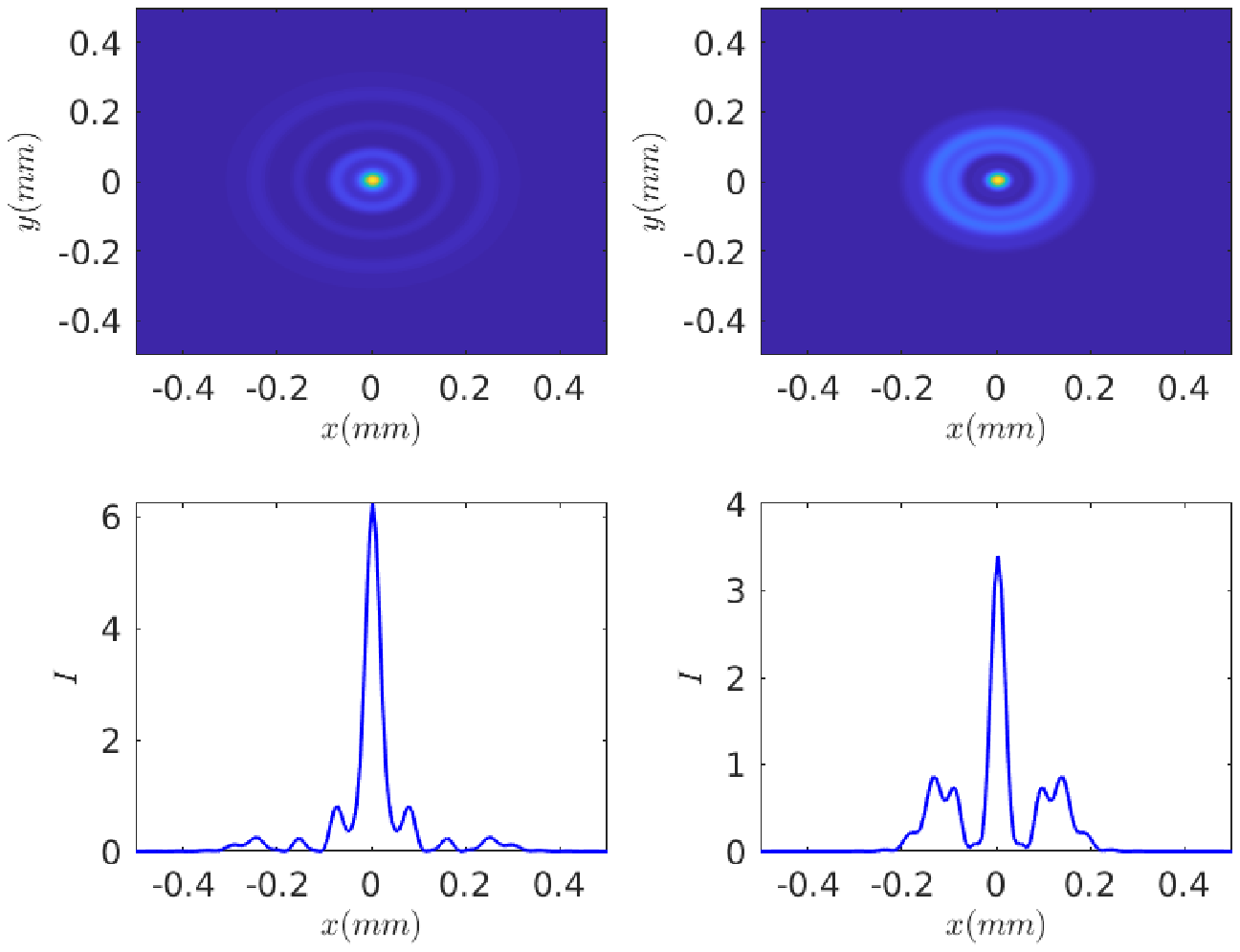}
\caption{Left plots: intensity profile predicted by the NSE model as a function of the two transverse spatial coordinates (top left) and $y=0$ slice (bottom left). Right plots: intensity profile predicted by the SM model as a function of the two transverse spatial coordinates (top right) and $y=0$ slice (bottom right). All plots are at P=5.5W, as in both cases the central peak is at an absolute maximum at that power.}
\label{NSEvsSM2}
\end{figure}

The figure \ref{NSEvsSM1} shows a quantitative comparison between 
the NSE and the SM models: the dynamics are qualitatively similar, 
showing in both cases a collapse followed by nonlinear oscillations. 
At low power, the difference between the structure of the output 
intensity profiles is quite small, while after the minimum of the 
$R(P)$ curve, the dynamics of the two systems start to differ 
significantly. In particular, one can see for the NSE an oscillating 
peak surrounded by concentric rings, while for the SM result the 
rings are less extended and the peak less pronounced (Fig.~\ref{NSEvsSM2}). Indeed, the NSE output intensity profile is in 
agreement with the one observed in the experiment, as at high power 
it is characterized by a central oscillating soliton surrounded by the 
classical solution, whereas 
the Snyder-Mitchell result is quite different, being more similar 
to a Fresnel diffraction pattern from a circular aperture.

\begin{figure}[H]
\centering
\includegraphics[scale=0.8]{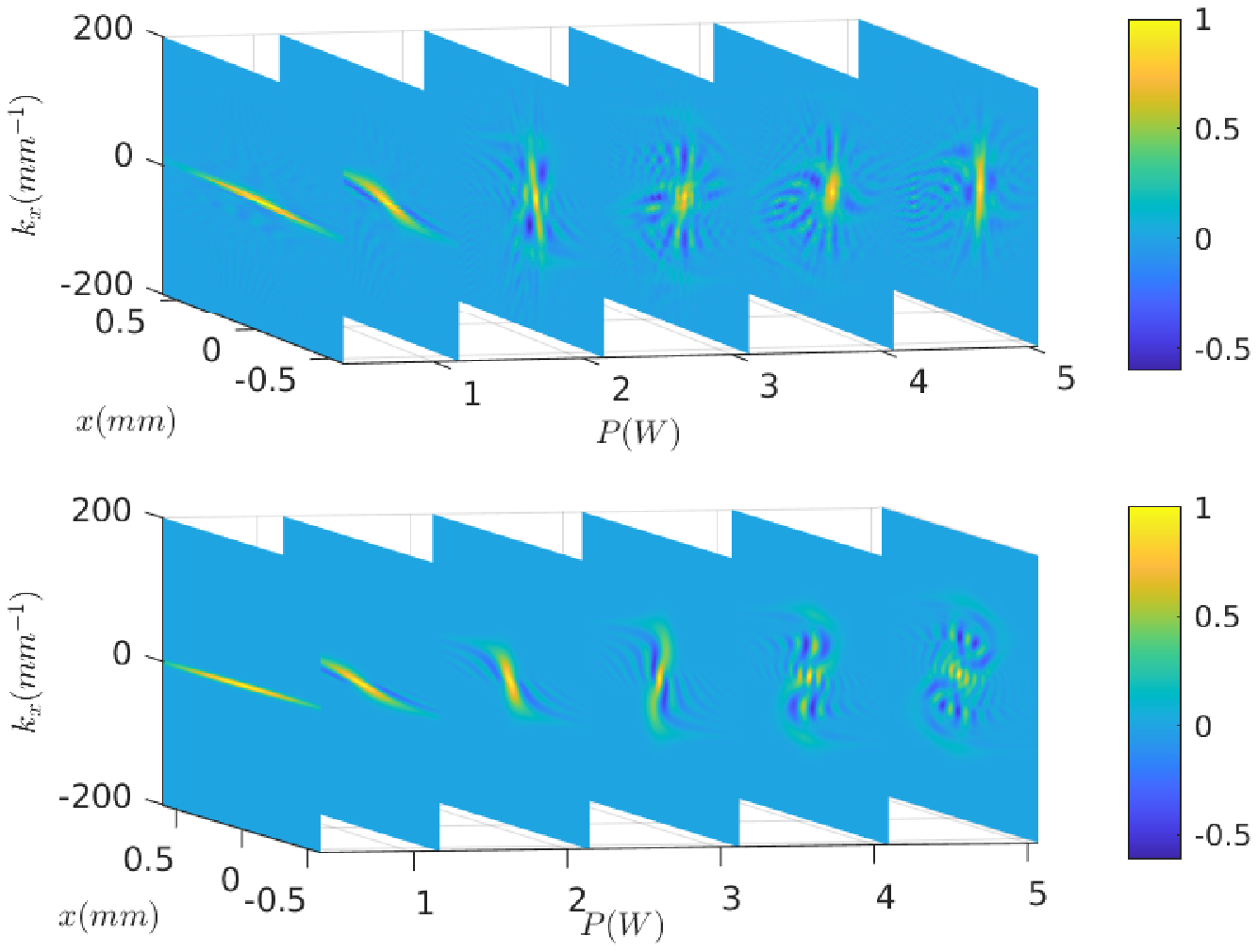}
\caption{Results of the NSE simulation (first row) and SM model (second row) for the $y = 0, k_y = 0$ profiles of the Wigner distribution.}
\label{NSEvsSM3}
\end{figure}

We arrive at the same conclusions looking at the phase-space 
dynamics, shown in Fig.~\ref{NSEvsSM3}: at the beginning the 
two models are very similar, while at high powers, and in 
particular after the collapse ($P\approx1.8\,\mathsf{W}$), 
the NSE model exhibits a more complicated dynamics (akin to 
the experimental Wigner distribution) compared with SM.

These comparisons allow to discriminate the static an-harmonic 
potential and represent a further confirmation of presence of 
violent relaxation in the experimental system. 

\section{Finite $\chi$ effects in the evolution of the energy distribution}
\label{finite-size-effects}

\begin{figure}[H]
\centering
 \includegraphics[width=\textwidth]{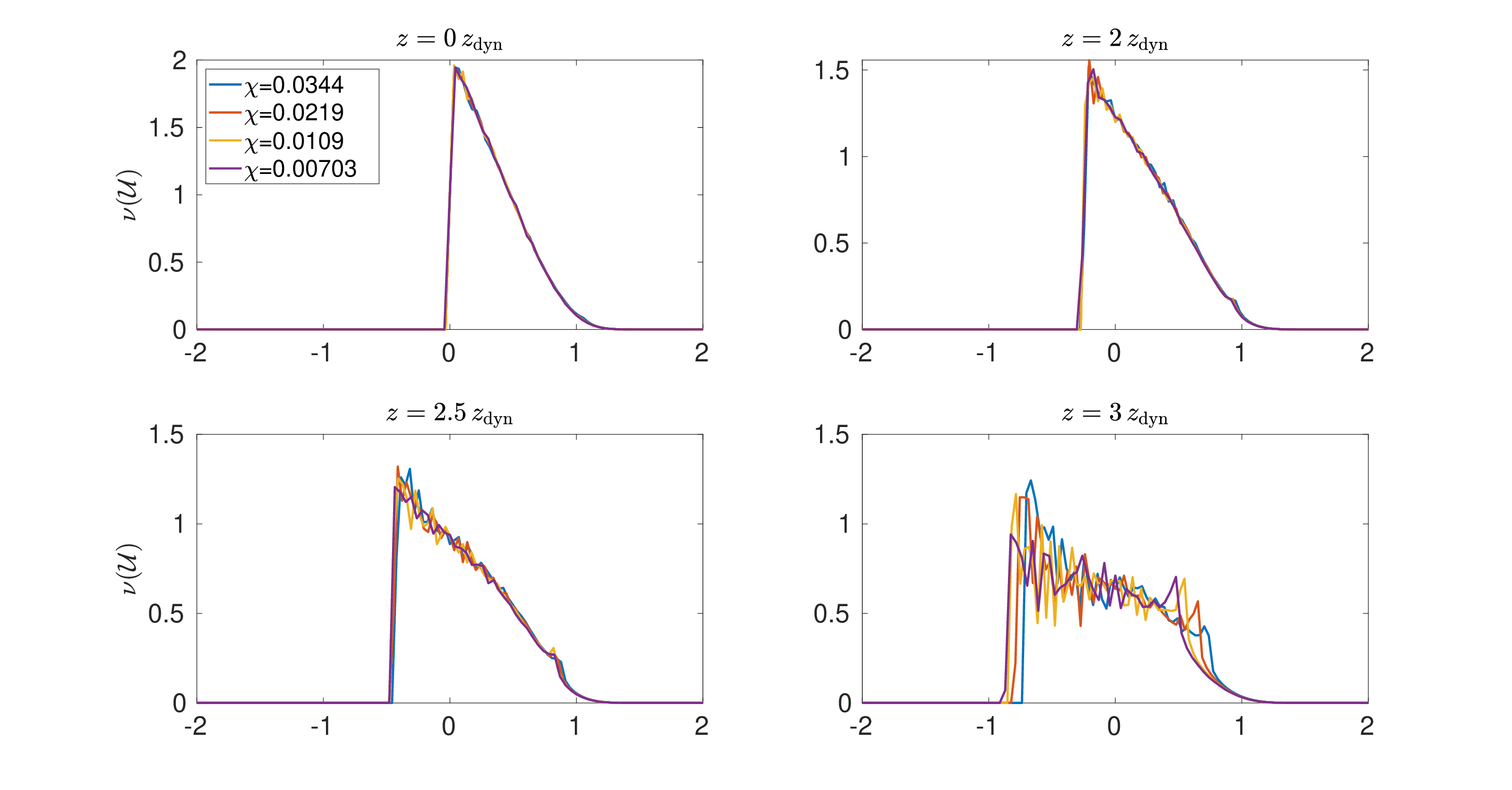} \\
\includegraphics[width=\textwidth]{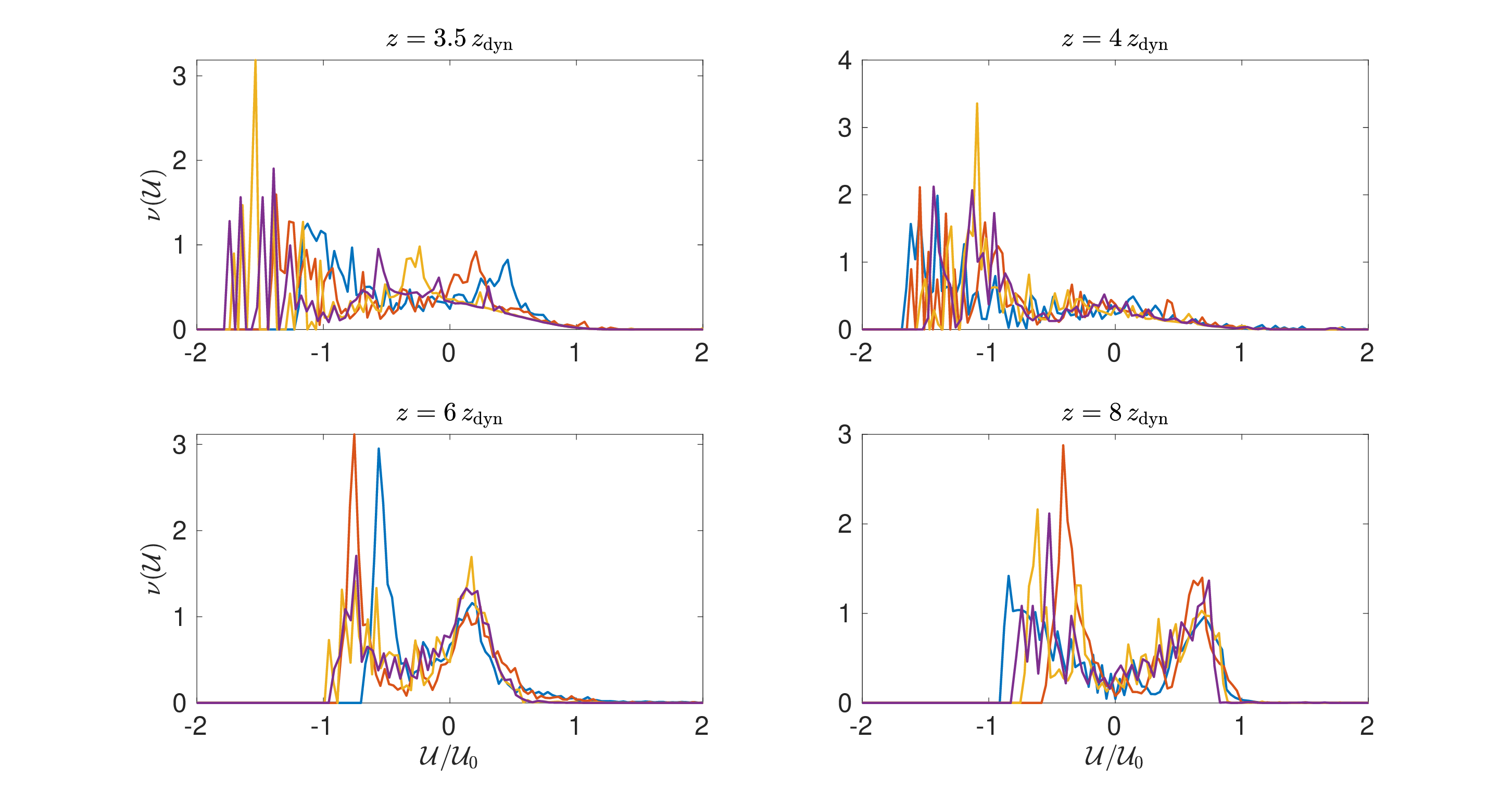}
\caption{Simulation of the evolution of the energy distribution for different values of the power $\chi$. The energy axis is in units of $\sE_0=\left(\alpha \beta k_0 P\right)/\left(2 \pi \kappa\right)$.}
\label{diffraction-effect}
\end{figure}

    We determine the importance of finite $\chi$ effects in the evolution of the energy distribution, performing simulations with different values of  $\chi$, without dissipation and reflections.   We plot the evolution of the energy distribution in  Fig.~\ref{diffraction-effect}, for different values of $z$, expressed in units of $z_{\mathrm{dyn}}$. In the experiment $z_{\mathrm{dyn}}=6.2\,cm$ for $P\approx 1.77\,W$ (see Tab.~\ref{table-chi2}). For values  $z\lesssim 3\,z_{\mathrm{dyn}}$ we observe a weak finite $\chi$ effect, for all values simulated. For $z\gtrsim 3\,z_{\mathrm{dyn}}$ we observe a convergence for the smallest values of $\chi$, which coincides with the corresponding values of $\chi$ used in the experiment. We can conclude that the experiment $\chi$ is sufficiently small in order to have little incidence on the change of the energy distribution compared with violent relaxation.

\begin{table}[h!]
\begin{center}
\begin{tabular}{ |c|c|c| }
	\hline
	$z/z_{\mathrm{dyn}}$ & P & $\chi$ \\
	\hline
	1.20 & 0.11 & 0.0680 \\
	1.84 & 0.26 & 0.0442 \\
	2.40 & 0.44 &  0.0340\\
	3.02 & 0.70 & 0.0269 \\
	3.61 & 1.00 &  0.0225\\
	4.81 & 1.77 &  0.0169\\
	7.21 & 3.98  &  0.0113\\
	8.32 & 5.30 &  0.0098\\
	\hline
\end{tabular}
\end{center}
\caption{Values of $z$ corresponding at the power $P$ in the experiment with its associated value of $\chi$.} 
\label{table-chi2}
\end{table}

\section{Losses effect}

In the absence of losses and in the limit $\chi\to 0$ the mapping between $P$ and $z$ used in the paper is perfect and the only mechanism responsible to the evolution of the energy distribution is violent relaxation. We have seen in Sect.~\ref{finite-size-effects} of Supp. Inf. that finite $\chi$ effects are small. We study here the relevance of losses. 

\begin{figure}[H]\label{losses-effect}
\includegraphics[width=\linewidth]{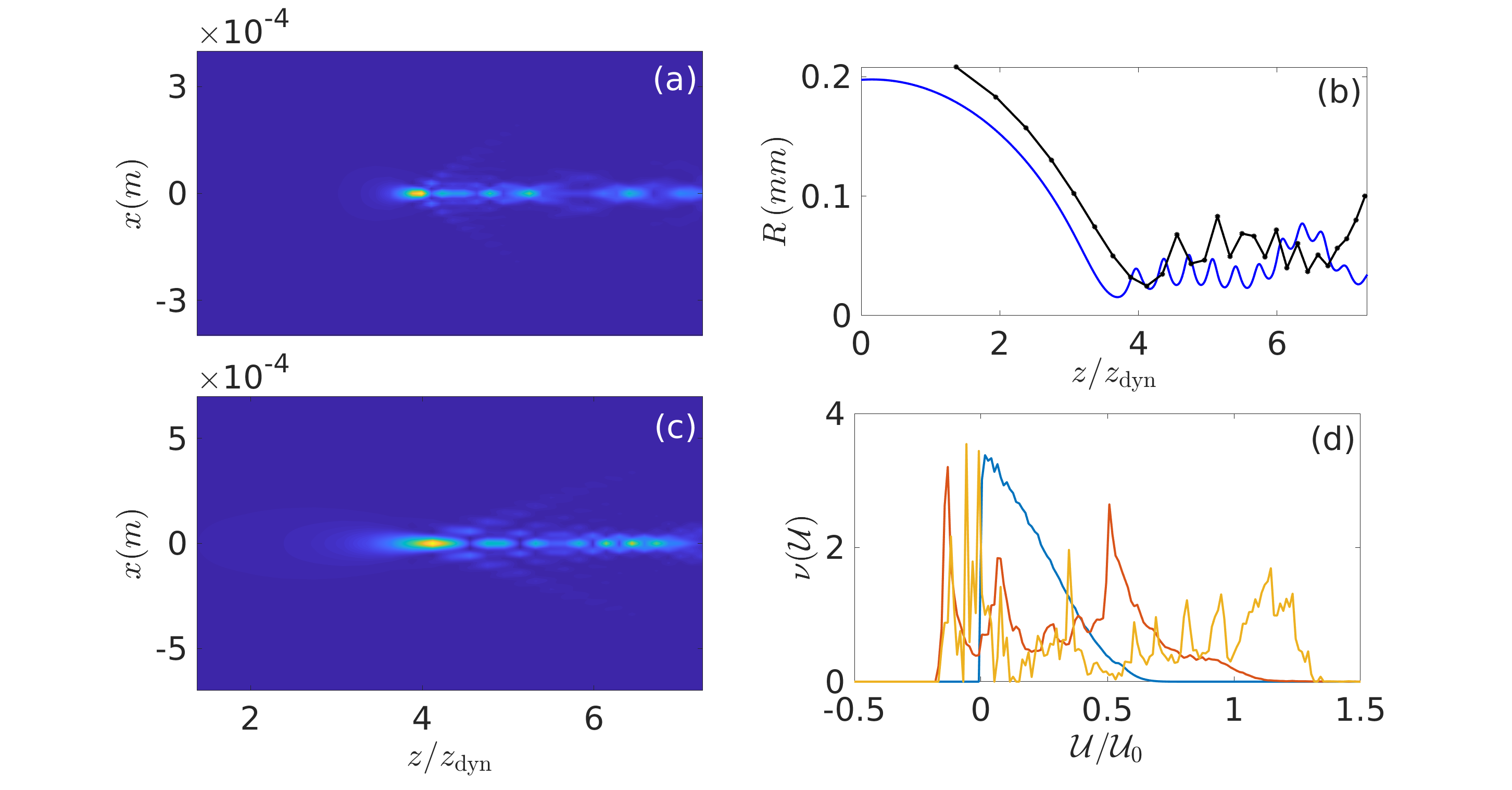}
\caption{(a): evolution of the intensity profile  $I(x,y=0,z;P=5.5\,W)$ for a simulation without losses. (b): Evolution of the transverse size of the beam $R(z)$ without losses (blue curve)  and with losses (black curve).
(c): evolution of the intensity profile  $I(x,y=0,z=30\,cm;P)$ for a simulation with losses, with $P$ expressed in terms of $z/z_\mathrm{dyn}$ (see text). (d): evolution of the energy distribution from the initial condition (blue curve) to the largest power, with losses (orange curve) and without losses (yellow curve).}
\end{figure}

 First we investigate the effect of losses in the mapping between $P$ and $z$. In Fig.~\ref{losses-effect}(a)-(c) we show the difference in the evolution of the intensity $I=|\mE|^2$ between studying the dynamics in $z$ (which corresponds to the original system) or in terms of the power (which is the way according to which the experiment has been performed, as it is impossible to measure the amplitude of the beam inside the material). Specifically, in Fig.~\ref{losses-effect}(a) we show the evolution of the beam intensity profile $I(x,y=0,z;P=5.5\,W)$ as a function of the propagation coordinate $z/z_\mathrm{dyn}$, at fixed power, obtained from a simulation without losses. In Fig.~\ref{losses-effect}(c) on the other hand, we show the evolution of intensity profile $I(x,y=0,z=L=30\,cm;P)$, obtained from a simulation with losses, varying the power, and expressing the propagation coordinated as $L/z_\mathrm{dyn}$. The latter reproduces the experimental configuration. In both simulations we observe the same qualitative behaviour. In Fig.~\ref{losses-effect}(b) we show the evolution of the transverse size of the system $R(z)$ for the same simulations presented in (a) and (c): without losses (blue curve) and with losses (black curve). The main difference is that the black curve collapses later compared with the blue one. This is due to the presence of losses, which slow-down the dynamics. For the same reason, the average beam size obtained varying the power at constant $z$, is shown to be slightly larger, compared to the blue curve. Despite these differences, the undergoing physics remains the same.

We investigate now the effect of losses in the evolution of the energy distribution after the collapse. In Fig.~\ref{losses-effect}(d) we show the evolution of the energy distribution, obtained from numerical simulations, from the lowest power to the largest one. The orange curve corresponds to the experimental configuration, with losses. The yellow curve correspond to a simulation without losses. In order to have both curves corresponding to the same output intensity, the yellow one is multiplied by a factor $T^5e^{-\alpha L}$, which take into account the losses. We observe that the difference form the initial energy distribution is much larger than the differences between these curves, which allows to conclude that the effect of violent relaxation dominates over losses in the evolution of the energy distribution.

\section{Evolution of the energy distribution for each power}

\begin{figure}
\centering
\includegraphics[width=\textwidth]{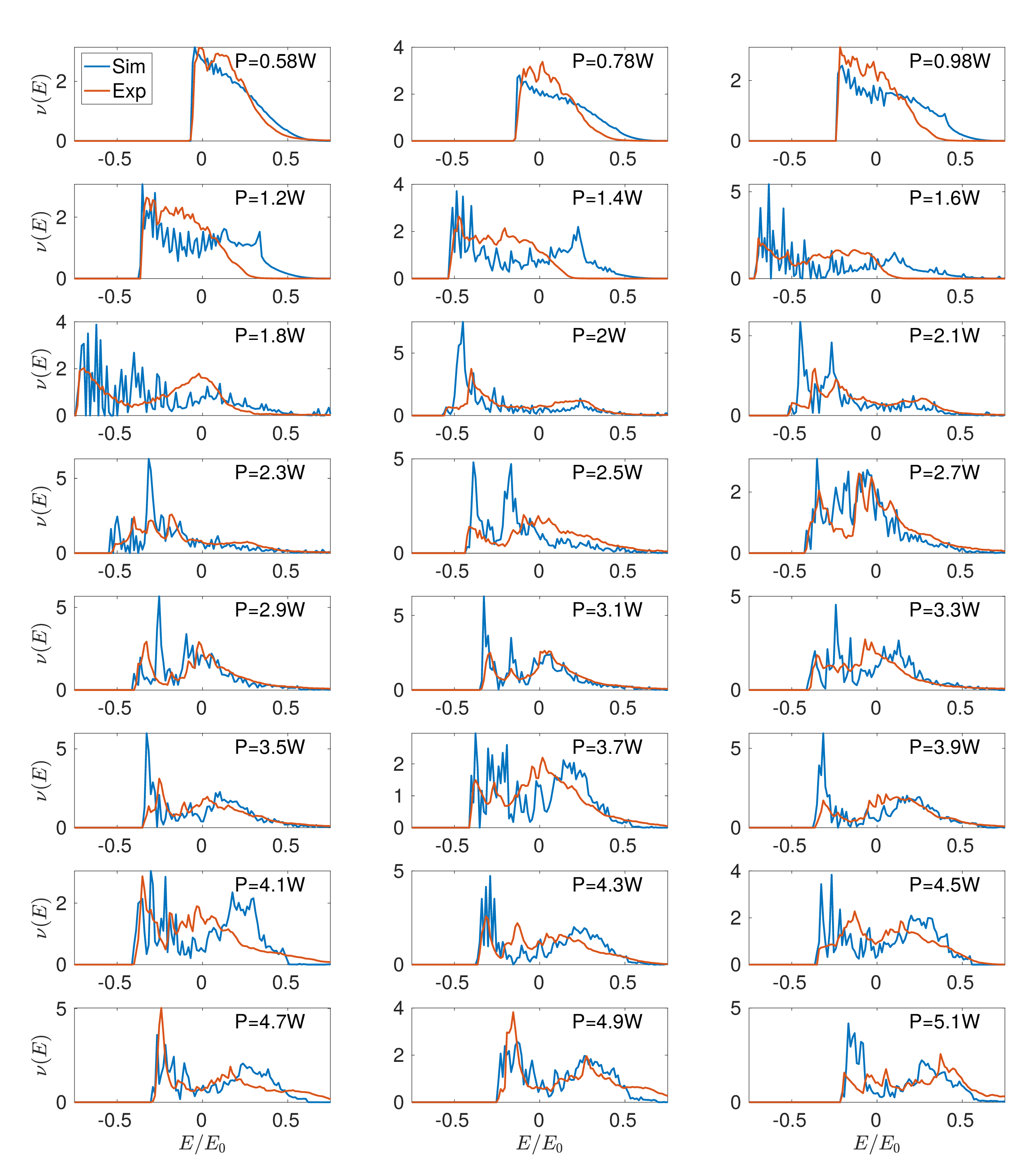}

	\caption{Energy distribution at different values of power for simulation (blue curves) and experiment (red lcurves). The energy axis is in units of $\sE_0=\left(\alpha \beta k_0 P\right)/\left(2 \pi \kappa\right)$.}
\label{energy_comparison2}
\end{figure}

In Fig.~\ref{energy_comparison2} we show the evolution of the energy distribution for each power. We observe a good quantitative agreement between experience (red curves) and simulation (blue curve). 
